\documentclass[a4paper,twocolumn,11pt,accepted=2026-06-03]{quantumarticle}
\pdfoutput=1

\usepackage[T1]{fontenc}
\usepackage{amsfonts}
\usepackage{amsmath}
\usepackage{amssymb}
\usepackage{amsthm}
\usepackage{bm}
\usepackage{mathtools}
\usepackage{dsfont}
\usepackage[dvipsnames]{xcolor}
\usepackage{enumitem}

\usepackage{caption}
\usepackage{multirow}
\usepackage{float}

\usepackage{arydshln}
\usepackage{cellspace}
\usepackage{microtype}
\usepackage{listings}
\usepackage{algorithm}
\usepackage{algpseudocode}

\newcommand{\iu}{\mathrm{i}}

\newcommand{\tr}{\mathrm{Tr}}
\newcommand{\fld}{\mathbb{Z}_d}

\usepackage{natbib}
\usepackage{hyperref}

\begin{document}

\title{Qudit Clauser-Horne-Shimony-Holt Inequality and Nonlocality from Wigner Negativity}

\author{Uta Isabella Meyer}
\author{Ivan Šupić}
\author{Damian Markham}
\author{Frédéric Grosshans}
\affiliation{Sorbonne Université, CNRS, LIP6, F-75005 Paris, France}

\begin{abstract}
Nonlocality is an essential concept that distinguishes quantum from classical models and has been extensively studied in systems of qubits.
For higher-dimensional systems, certain results for their two-level counterpart, like Bell violations with stabilizer states and Clifford operators, do not generalize.
On the other hand, similar to continuous variable systems, Wigner negativity is necessary for nonlocality in qudit systems.
We propose a new generalization of the CHSH inequality for qudits by inquiring correlations related to the Wigner negativity of stabilizer states under the adjoint action of a generalization of the qubit $\pi/8$-gate.
A specified stabilizer state maximally violates the inequality among all qudit states based on its Wigner negativity.
The Bell operator not only serves as a measure for the singlet fraction but also quantifies the volume of Wigner negativity.
Additionally, we show how a bipartite entangled qudit state can serve as a witness for contextuality when it exhibits Wigner negativity.
Furthermore, we identify rational-phase diagonal unitaries as the key resource that exactly reproduce the CGLMP and SATWAP violation with the maximally entangled state through simple phase-difference alignment.
\end{abstract}
\maketitle

\section{Introduction}
As a canonical extension of the classical bit as a unit of information, quantum information theory has extensively studied two-level systems, now called qubits.
A generalization to higher dimensions, systems of $d$ levels, are called qu\textit{d}its.
Although qubits are typically easier to describe, qudits can be more efficient for some quantum information tasks:
apart from the theoretical interest in deriving no-go theorems in quantum information~\cite{shaari2008blind,PhysRevA.78.012353,PhysRevLett.102.120501}, qudit-based algorithms for Byzantine agreement~\cite{Fitzi2001} and secret sharing in graph states~\cite{Markham2008,Kashefi_2009,Keet2010} appear to surpass schemes known for qubits~\cite{Cleve1999}.
Beyond any advantage over qubit systems, quantum information with qudits broadens the range of physical platforms where quantum information tasks can be implemented, as many quantum systems are naturally higher dimensional.

We study Bell nonlocality in systems of two or more qudits, i.e.\ correlations that cannot be reproduced by local-hidden-variable models~\cite{bell1964einstein,brunner2014bell}. 
Nonlocality in high-dimensional systems has been investigated using Bell states~\cite{Collins2001,Liang2009,Ji2008,salavrakos2017bell}, GHZ states~\cite{Cerf2001}, and graph states~\cite{Tang2013}. 
However, many proposed inequalities are technically involved and difficult to analyze, and it is often unclear to what extent they witness genuinely $d$-dimensional effects rather than properties inherited from subsystems (e.g.\ by restricting to subsets of levels). 
For $d=3$, intrinsic approaches and self-testing results are known~\cite{Lawrence2017,Kaszlikowski2002,Kaszlikowski2002-2,Acin2004,Gruca2011,Li2011,Mackeprang2023,Kaniewski2018}. 
In a complementary direction, Howard~\cite{Howard2015bell} connects nonlocality to magic-state resources using tools from~\cite{Howard2012}, which we also employ here.

Our main contribution is a qudit generalization of the qubit Clauser-Horne-Shimony-Holt (CHSH) construction that makes the underlying mechanism transparent and yields Bell operators that are easy to analyze. 
The key organizing principle is Gross' discrete Wigner function: in odd prime dimensions, Wigner negativity is equivalent to contextuality for projective Pauli measurements~\cite{Gross2006,Howard2014,Delfosse2017}.
Thus Wigner negativity, in the state or effectively in the measurements, is a necessary ingredient for Bell nonlocality (similarly to continuous variables~\cite{PhysRevLett.129.230401}). 
We implement this idea by constructing a family of Heisenberg-Weyl (HW) Bell operators with a direct phase-space interpretation.

\paragraph*{Summary of results.}
\begin{enumerate}
    \item \textbf{Phase-space Bell operators from characteristic functions.}
    We introduce a family of qudit Bell operators whose coefficients are given by the characteristic function of a (possibly rotated) entangled state, leading to the state-dependent construction in Eq.\,\eqref{eq:belloperatorgen} and the rotated-Bell-state family in Eq.\,\eqref{eq:belloperatordisplace}.

    \item \textbf{Contextuality and phase-space bounds.}
    Using phase-space methods in odd prime dimension, we derive non-contextual bounds and connect them to constraints on the discrete Wigner function, including the negativity-volume bound Eq.\,\eqref{eq:volumebound}.

    \item \textbf{Nonlocality bounds and Bell violations.}
    We derive the corresponding LHV bound in the HW setting (Eq.\,\eqref{eq:bellineq}) and show how violations arise in the presence of Wigner negativity, with the Bell expectation admitting a singlet-fraction interpretation within our construction.

    \item \textbf{Reducing settings and extending the construction.}
    We show how stabilizer symmetries and the diagonal structure of the rotations reduce the number of required measurement settings, and we extend the framework to multipartite stabilizer states as well as to other classes of states.

    \item \textbf{Connections to other high-dimensional inequalities and resources.}
    We relate the rotated-HW viewpoint to the CGLMP inequality and the SATWAP family (Sec.\,\ref{sec:otherinequalities}), and we identify rational-phase diagonal unitaries as key non-Clifford resources generating the relevant maximally entangled correlators.

    \item \textbf{Concrete example family and numerics.}
    For unitary cube operators we provide analytic control of non-contextual bounds and compute LHV bounds numerically (Sec.\,\ref{sec:numerics}), with technical details and the optimization procedure given in the appendices.
\end{enumerate}

\section{Preliminaries}

We restrict our study to odd prime dimensions $d$, so that $\mathbb{Z}_d$ forms a finite field and all arithmetic is taken modulo $d$. In particular, $2^{-1}$ is well-defined and unambiguous. This choice also excludes features that rely on a nontrivial factorization of the local dimension (which can arise for composite $d$).

The qudit operators $X$ and $Z$ are natural generalizations of the qubit Pauli operators $\sigma_x$ and $\sigma_z$.
They act on the computational basis as
$X \vert k \rangle = \vert (k+1) \mathrm{mod}\,d \rangle$ and $Z \vert k \rangle = \omega^k \vert k \rangle $, where $\omega=\exp(2\pi \mathrm{i}/d)$ is a $d^{\text{th}}$ root of unity.
For $x,z\in\fld$, they satisfy $Z^z X^x = \omega^{xz} X^x Z^z $, and are unitary with $X^{-1}=X^{d-1}=X^\dagger$ and $Z^{-1}=Z^{d-1}=Z^\dagger$.
Then, the qudit (Heisenberg-Weyl) displacement operators are
\begin{equation}
T_{(x,z)} = \omega^{2^{-1}xz} X^x Z^z \,.
\label{eq:displace}
\end{equation}
For any $v\in\fld^2$, $T_v$ has eigenvalues labeled by $m\in\fld$ and spectral projectors
\[
\Pi^{(v)}_m = \frac{1}{d}\sum_{k\in\fld}\omega^{mk} T_v^k \,,
\]
which define the corresponding projective measurement in the eigenbasis of $T_v$.

The qudit Pauli group $\mathcal{P}$ decomposes into $d+1$ subgroups of commuting operators:\@ for $r=0,\dots,d-1$,
\[ \mathcal{G}_r =\{T_{(s,sr)}\,:\,s\in\fld\} \,, \quad \mathcal{G}_d =\{T_{(0,z)}\,:\,z\in\fld\} \,. \]
Any qudit state has the unique expansion
$\rho=\sum_{u\in\fld^2}\chi_u[\rho]\,T_u/d$ in terms of the characteristic function
\begin{equation}
\chi_u[\rho]=\tr(\rho\,T_{-u})\,.
\label{eq:characteristic_function}
\end{equation}
This extends straightforwardly to multiple qudits; for instance, if $\rho$ is a two-qudit state then $\chi_{u,v}[\rho]=\tr(\rho\,T_{-u}\otimes T_{-v})$\,.

The discrete Fourier transform of the characteristic function yields Gross's Wigner function~\cite{Gross2006}, a real-valued, normalized quasi-probability distribution that may take negative values.
Writing $u=(u_x,u_z)\in\fld^2$ and using the symplectic form $[u,v]:=u_z v_x-u_x v_z$, we have
\begin{equation}
W_u[\rho]=\frac{1}{d}\sum_{v\in\fld^2}\omega^{-[u,v]}\chi_v[\rho]\,.
\label{eq:wigner}
\end{equation}
The negativity volume is defined as the sum of all negative terms or
\begin{equation}
    N[\rho] = \frac{1}{2} \Big( \sum_{u \in \fld^2} \left\vert W_u[\rho] \right\vert - 1 \Big)
\end{equation}

For a system of $n$ qudits, we write $\bm u=(\bm u_x,\bm u_z)\in\fld^{2n}$ and define $T_{\bm u}=\bigotimes_{i=1}^n T_{((u_i)_x,(u_i)_z)}$, which in turn defines $\chi_{\bm u}[\rho]$ and $W_{\bm u}[\rho]$ analogously for $n$-qudit states $\rho$.
We also use the standard inner product on $\fld^n$,
\[
\bm u\cdot \bm v=\sum_{i=1}^n u_i v_i\,.
\]
so that the symplectic form on $\fld^{2n}$ can be written compactly as
\[
[\bm u,\bm v]=\bm u_z\cdot \bm v_x-\bm u_x\cdot \bm v_z\,.
\]

To study nonlocality, we start with the bipartite Bell scenario for two qudits.
Alice chooses a local setting $x$ and Bob chooses a local setting $y$.
They obtain outcomes $a$ and $b$, and the experiment is described by the conditional probabilities $p(a,b\vert x,y)$.

In quantum theory, these probabilities arise from a bipartite state $\rho$ and local projective measurements $\{\Pi_{a\vert x}\}_a$ and $\{\Pi_{b\vert y}\}_b$ via Born rule,
\[
p(a,b\vert x,y)=\tr\left[(\Pi_{a\vert x}\otimes \Pi_{b\vert y})\,\rho\right].
\]
A (classical) local-hidden-variable (LHV) model is more restrictive: the correlations must admit a decomposition
\[
p(a,b\vert x,y)=\sum_{\lambda}\mu(\lambda)\,p_A(a\vert x,\lambda)\,p_B(b\vert y,\lambda),
\]
for some distribution $\mu(\lambda)$ and local response functions $p_A,p_B$.
Throughout, outcomes are labeled by $a,b \in \fld$ and encoded as phases $\omega^a,\omega^b$ when forming correlators.
This is a natural extension from the qubit case ($d=2$) where $\omega = -1$ and $a,b \in \{ 0,1 \} \rightarrow \omega^a,\omega^b \in \{ \pm 1\}$.

To witness a separation between quantum and LHV correlations, we consider Bell inequalities built from observables $O_x$ and $O_y$ with eigenvalues taken as the outcomes.
With coefficients $c_{xy}$, define the Bell operator
\[
\mathcal B=\sum_{x,y} c_{xy}\, O_x\otimes O_y.
\]
For an LHV model,
\begin{equation}
\langle \mathcal B\rangle_{\mathrm{LHV}}
=\!\sum_{a,b,x,y,\lambda}\!\mu(\lambda)\,c_{xy}\,\omega^{a+b}\,p_A(a\vert x,\lambda)\,p_B(b\vert y,\lambda),
\label{eq:lhvgeneral}
\end{equation}
and the corresponding local bound is $\langle \mathcal B \rangle_{\mathrm{LHV}} \leq \max \langle \mathcal B\rangle_{\mathrm{LHV}} =: B_{\mathrm{LHV}}$.
Thus any observed distribution for which
\[
\sum_{a,b,x,y} c_{xy} \, \omega^{a+b} \,p(a,b\vert x,y) > B_{\mathrm{LHV}}
\]
cannot be explained by a local-hidden-variable model. Throughout this work, the measurements are projective with non-degenerate spectra.

\section{A new qudit CHSH operator.}
A notorious bipartite Bell inequality for experiments with two measurement outcomes is the CHSH inequality, which is maximally violated by local Pauli measurements (a subset of Clifford measurements) on two qubits prepared in a Bell state. The corresponding Bell operator can be written as
\begin{equation}
    \mathcal{B}_2 = \sum_{i,j=0,1} (-1)^{ij} A_i \otimes B_j \,.
    \label{eq:belloperatorqubit}
\end{equation}
A natural generalization of this inequality to qudits replaces the second root of unity $-1$ by the $d^{\text{th}}$ root of unity $\omega=\exp(2\pi \iu/d)$, leading to
\begin{equation}
    \mathcal{B}_d = \sum_{n,i,j=0}^{d-1} \omega^{n i j} A_i^n \otimes B_j^n \,.
    \label{eq:belloperatorold}
\end{equation}
The sum over $k$ ensures that $\mathcal{B}_d$ is Hermitian. Operators of this type and closely related constructions have been studied extensively~\cite{Liang2009,Ji2008,salavrakos2017bell,Kaniewski2018,Howard2015bell}, and can witness nonlocality either using local qudit Clifford measurements on a qudit Bell state, or equivalently using Pauli measurements on a Bell state rotated by suitable local (possibly non-Clifford) unitaries.

Our starting point is the observation that the qubit operator~\eqref{eq:belloperatorqubit} can also be written using the characteristic function of a (rotated) Bell state,
\begin{equation}
    \mathcal{B}_2 = \sum_{u,v=(0,1),(1,0)} \chi_{u,v}[\vert \Phi_{\pi/8} \rangle \langle \Phi_{\pi/8} \vert] \,A_u \otimes B_v \,,
    \label{eq:qubitbellcharacter}
\end{equation}
where for bipartite qubit states $\chi_{u,v}[\rho]=\tr(\rho\,\sigma_u\otimes\sigma_v)$ and
$\vert \Phi_{\pi/8} \rangle = (\mathds{1} \otimes R_{\pi/8})(\vert 00\rangle+\vert 11\rangle)/\sqrt{2}$ with
$R_{\pi/8}=\exp(\iu\pi\sigma_z/8)$.
Extending the sum in~\eqref{eq:qubitbellcharacter} to all $u,v\in\mathbb{Z}_2^2$ adds two further non-zero terms which can be treated independently and does not change the achievable nonlocal violation (see Appendix~\ref{app:chsh}).

\paragraph*{A qudit family from state-dependent coefficients.}
Motivated by~\eqref{eq:qubitbellcharacter}, we define a broad family of qudit Bell operators by taking the coefficients from the characteristic function of a chosen bipartite qudit state $\rho$:
\begin{equation}
    \mathcal{B}[\sigma] = \frac{1}{d^2} \sum_{u,v\in\fld^2} \chi_{u,v}[\sigma] \, A_u \otimes B_v \,.
    \label{eq:belloperatorgen}
\end{equation}
A particularly relevant choice, mimicking the qubit case, is a rotated qudit Bell state $\sigma=\vert \Phi_U \rangle \langle \Phi_U \vert$ with $\vert \Phi_U \rangle = (\mathds{1}\otimes U) \sum_{j\in\mathbb{Z}_d} \vert jj \rangle/\sqrt{d} $
for some unitary $U$.

Although both~\eqref{eq:belloperatorold} and~\eqref{eq:belloperatorgen} may be viewed as qudit extensions of~\eqref{eq:belloperatorqubit}, they define intrinsically different Bell scenarios.
The operator~\eqref{eq:belloperatorold} uses $d$ measurement settings per party, whereas~\eqref{eq:belloperatorgen} is naturally indexed by $u,v\in\fld^2$ (hence $d^2$ labels per party before exploiting structure).
In particular,~\eqref{eq:belloperatorgen} defines a large class of Bell operators parameterized by (essentially) any bipartite state $\sigma$.
We will show that taking $\sigma$ to be a Bell state rotated by suitable non-Clifford unitaries leads to inequalities that are maximally violated by the corresponding rotated Bell state.

\paragraph*{Two equivalent quantum realizations.}
In the remainder of this work we use two equivalent ways to realize~\eqref{eq:belloperatorgen} in quantum theory:
(i) the coefficients are taken from a (possibly non-stabilizer) state $\sigma$ and the measurements are Pauli measurements; or
(ii) the shared state is the standard qudit Bell state and the measurements are related Clifford measurements.
We later exploit (ii) to construct variants with fewer measurements and variants tailored to general qudit stabilizer states.

\section{Contextuality and nonlocality bounds}

On at least two qudits, a non-contextual (NC) model describes the outcomes of Pauli measurements corresponding to $T_{u_1} \otimes T_{u_2}$ (if they can be performed simultaneously) as multiplicative characters of the form $\omega^{[a_1,u_1]+ [a_2,u_2]}$ for $a_1,a_2 \in \fld^2$~\cite{Delfosse2017}.
Such a deterministic non-contextual value assignment (NC$^\ast$) can only achieve
\begin{align}
\langle \mathcal{B}[\sigma] \rangle_{\mathrm{NC}^{\ast}} &= \frac{1}{d^2}\sum_{u_{1},u_{2} \in \fld^2} \chi_{u_{1},u_{2}}[\sigma]\, \omega^{[a_1,{u_{1}}]} \omega^{[a_2,{u_{2}}]}\nonumber \\
&= d^2 \,W_{a_1,a_2}[\sigma]\,. \label{eq:belloperatorlhv}
\end{align}
As a result, any NC model, a convex mixture of deterministic classical value assignment (NC$^\ast$), can maximally achieve
\begin{equation} \left\langle \mathcal{B}[\sigma] \right\rangle_{\mathrm{NC}} \leq d^2 \max_{a_1,a_2 \in \fld^{2}} \vert W_{a_1,a_2} [\sigma] \vert \equiv B_{\mathrm{NC}}^{\max} \,. \label{eq:ncineq} \end{equation}
On the other hand, the pure state $\rho = \sigma$ maximally violates the non-contextual inequality, which it defines via its characteristic function, among all qudit states if $\tr (\rho^2) = 1 > B_{\mathrm{NC}}^{\max}$.
We verify that non-trivial negativity volume, $N[\Psi]>0$, is necessary for such a violation, see Eq.\,\eqref{eq:wignerbounds}.
As a result, Eq.\,\eqref{eq:belloperatorgen} defines a family of inequalities based on Wigner negative states that are witnesses of contextuality if $1 > d^2 \max_{a_1,a_2 \in \fld^{2}} \vert W_{a_1,a_2} [\rho]\vert $.
This specifically exploit the equivalence of Wigner negativity and contextuality with Pauli measurements~\cite{Delfosse2017}.

If we spatially separate the two qudits, the Pauli measurements on Alice's and Bob's side cannot be performed jointly, and a local-hidden-variable (LHV) model is less restrictive than a non-contextual one.
In particular, a deterministic LHV strategy assigns a phase to each of the $d+1$ local measurement settings (i.e., to each commuting Pauli subgroup) independently on the two parties.
Writing $r(u)\in\{0,\dots,d\}$ for the label of the commuting subgroup containing $u\in\fld^2$ and fixing the identity value $T_0=\mathds{1}$ to $1$ (equivalently $\alpha_0=\beta_0=0$), we obtain the LHV bound
\begin{equation}
B^{\max}_{\mathrm{LHV}}
:= \frac{1}{d^2}\max_{\alpha,\beta\in\fld^{d+1}}
\sum_{u_1,u_2\in\fld^2}\chi_{u_1,u_2}[\sigma]\,
\omega^{\alpha_{r(u_1)}+\beta_{r(u_2)}} \,.
\label{eq:bellineq}
\end{equation}
To observe a Bell violation, it is necessary that $B^{\max}_{\mathrm{LHV}} < 1$.
In general, evaluating $B^{\max}_{\mathrm{LHV}}$ requires either an explicit (often brute-force) search over deterministic strategies (which is feasible in the prime dimensions considered here), or additional analytic input about the structure of $\chi[\rho]$ (equivalently the Wigner function) that can be exploited to obtain sharper bounds.
We use both approaches: exhaustive optimization for the dimensions studied numerically, and analytic bounds in structured special cases.

\section{Main construction}

Qudit stabilizer states do not exhibit Wigner negativity.
They form a large class including Bell states, GHZ states, and graph states, and each stabilizer state $\vert S\rangle$ is defined by an Abelian subgroup $\mathcal{S}$ of the Pauli group.
Their (Gross) Wigner function is non-negative and remains so under Clifford operations, since the Clifford group normalizes the Pauli group.
This motivates the following realization of the Bell operator~\eqref{eq:belloperatorgen} which mirrors the qubit CHSH construction: we keep the (stabilizer) Bell state fixed and move the non-Clifford element into the measurement operators.

Let $\vert\Phi\rangle=\sum_{k=0}^{d-1}\vert k\,k\rangle/\sqrt d$ be the qudit Bell state and define the rotated Bell state $\vert\Phi_U\rangle := (U\otimes\mathds{1})\vert\Phi\rangle$.
We consider the Bell operator
\begin{equation}
\mathcal{B}_{U} = \frac{1}{d^2}\sum_{u_{1},u_{2} \in \fld^2} \chi_{u_{1},u_{2}}[\vert \Phi_U \rangle \langle \Phi_U \vert]~ U \,T_{u_{1}} U^\dagger \otimes T_{u_{2}} \,.
\label{eq:belloperatordisplace}
\end{equation}
By orthogonality of the Pauli basis, this is precisely the Pauli expansion of the Bell projector $\vert\Phi\rangle\langle\Phi\vert$ in the local operator basis
$\{U T_{u_1}U^\dagger\otimes T_{u_2}\}_{u_1,u_2}$.
Hence the Bell state attains the algebraic value
\[
\tr \!\left(\mathcal{B}_{U}\,\vert\Phi\rangle\langle\Phi\vert\right)=1,
\]
and, moreover, Eq.\,\eqref{eq:appsingletfraction} shows that for any bipartite state $\rho$ the expectation value is the singlet fraction
\begin{equation}
\langle \mathcal{B}_{U} \rangle_{\rho} = \tr \left( \mathcal{B}_{U} \rho \right) = \langle \Phi \vert \rho \vert \Phi \rangle \,.
\end{equation}

The non-contextual (NC) bound arises as in Eq.\,\eqref{eq:ncineq}: deterministic NC value assignments correspond to phase-space points and yield
$\langle\mathcal{B}_U\rangle_{\mathrm{NC}^\ast}=d^2\,W^{U}_{v_1,v_2}$.
Therefore,
\[
\left\langle \mathcal{B}_{U} \right\rangle_{\mathrm{NC}}
\leq d^2 \max_{v_1,v_2 \in \fld^{2}} \vert W_{v_1,v_2}^{U} \vert
\equiv B_{\mathrm{NC}}^{\max}\,.
\]
Finally, Eq.\,\eqref{eq:appcalcvolumebound} bounds the expectation value for any $\rho$ in terms of $B_{\mathrm{NC}}^{\max}$ and the negativity volume via H\"older's inequality,
\begin{equation}
\tr \left( \mathcal{B}_{U} \rho \right) \leq B_{\mathrm{NC}}^{\max} \bigl( 1 + 2N [ \vert \Phi_U \rangle \langle \Phi_U \vert ] \bigr)\,.
\label{eq:volumebound}
\end{equation}
In particular,
\( 1 \leq B_{\mathrm{NC}}^{\max} \bigl( 1 + 2N[\vert \Phi_{U} \rangle \langle \Phi_{U} \vert] \bigr)\,.\)
Thus, a contextuality (NC) violation with $\mathcal{B}_U$ requires a nontrivial negativity volume $N[ \vert \Phi_{U} \rangle \langle \Phi_{U} \vert ] > 0$.

\paragraph*{Reducing measurement settings via symmetries.}
The full Bell operator~\eqref{eq:belloperatordisplace} can be implemented using $(d+1)^2$ local Pauli measurement bases (one basis for each commuting subgroup on each party).
In practice, this is often redundant: one may reduce the number of settings by exploiting symmetries of the target state and of the chosen local unitary.
For instance, if the unitary operator $U$ is diagonal, it commutes with all $Z$-type Paulis.
Moreover, the Bell state $\vert\Phi\rangle$ is stabilized by the subgroup $T_{(x,z)}\otimes T_{(x,-z)}$, i.e., $T_{(x,z)}\otimes T_{(x,-z)}\vert\Phi\rangle=\vert\Phi\rangle$.
As a result, correlators stemming purely from this $Z\otimes Z^{-1}$ symmetry sector are fixed by the stabilizer and do not contribute to contextuality/nonlocality witnesses.
The nontrivial content comes from the remaining settings, in particular those involving the non-Clifford rotation on one subsystem.
Rather than introducing a separate reduced bipartite operator, we incorporate this symmetry restriction directly in the multipartite construction below, where the sum runs only over stabilizer labels and a single $Z$-shift parameter on the rotated subsystem.

\paragraph*{Extension to multipartite stabilizer states.}
Let $\vert S \rangle$ be an $n$-qudit stabilizer state with stabilizer group $\mathcal S\subset\mathcal P^{\otimes n}$.
We choose a set of labels ${\bm{u}}\in\Sigma\subset\fld^{2n}$ such that
$\omega^{t({\bm{u}})}T_{\bm{u}}\in\mathcal S$ for a linear function $t$; equivalently one may write
$t({\bm{u}})=[{\bm{a}},{\bm{u}}]$ for some ${\bm{a}}\in\fld^{2n}$.
Then $\omega^{t({\bm{u}})}T_{\bm{u}}\vert S\rangle=\vert S\rangle$ for all ${\bm{u}}\in\Sigma$, whereas for any Pauli operator not in the stabilizer the commutation relations imply $\langle S\vert T_{\bm{v}}\vert S\rangle=0$.

We now apply a unitary diagonal operator $U$ on the first qudit only and define $\vert S_\nu \rangle := (U \otimes \mathds{1}^{\otimes n-1})\vert S\rangle$.
Denoting the corresponding stabilizer-restricted characteristic coefficients by $\chi^{\nu}_{\bm{u}} := \langle S_\nu \vert T_{-{\bm{u}}}\vert S_\nu \rangle$ (restricted to ${\bm{u}}\in\Sigma$),
we consider the Bell operator
\begin{equation}\mathcal{B}_S = \sum_{{\bm{u}} \in \Sigma, t \in \fld } \chi^{\nu}_{{\bm{u}}} ~~ U \,T_{u_1+(0,t)} \, U^{\dagger} \bigotimes_{i = 2}^{n} T_{u_i} \,.
\label{eq:bellstab}
\end{equation}
This construction uses only stabilizer labels ${\bm{u}}\in\Sigma$ and a single $Z$-shift $t$ on the rotated subsystem, reflecting the symmetry reduction discussed above.

The operator $\mathcal{B}_S$ measures the overlap with $\vert S\rangle$ under the condition
$\bigl((0,\fld)_1\otimes(0,0)^{\otimes n-1}\bigr) \subset \Sigma$, which holds whenever $\vert S\rangle$ is entangled across the cut separating the first qudit from the rest.
In that case $\langle S\vert\mathcal{B}_S\vert S\rangle=1$, while the non-contextual bound satisfies
$\langle \mathcal{B}_S \rangle_{\mathrm{NC}} \le d^n \max_{{\bm{u}} \in \Sigma} W_{{\bm{u}}}^{\nu}<1$
by the same reasoning as in the bipartite case.
Finally, note that a single operator of the form~\eqref{eq:bellstab} cannot detect genuine multipartite entanglement beyond three parties~\cite{GUHNE20091}.

\paragraph*{Extension to other states}

The Bell operator $\mathcal{B}_U$ is tailored to the maximally entangled Bell state $\vert \Phi \rangle$, but one can equivalently consider one that is tailored to the rotated Bell state $\vert \Phi_U \rangle$,
\begin{align}
\nonumber &\mathcal{B}[\vert \Phi_U \rangle \langle \Phi_U \vert] \\ &= \frac{1}{d^2}\sum_{u_{1},u_{2} \in \fld^2} \chi_{u_{1},u_{2}}[\vert \Phi_U \rangle \langle \Phi_U \vert]\; T_{u_{1}}\otimes T_{u_{2}} \,.
\label{eq:belloperatordisplace2}
\end{align}
Their joint correlations are the exact same as those of $\vert \Phi \rangle$ and $\mathcal{B}_U$.
Naturally, any state $\sigma$ defines an operator
\begin{equation}
\mathcal{B}[\sigma] = \frac{1}{d^2}\sum_{u_{1},u_{2} \in \fld^2} \chi_{u_{1},u_{2}}[\sigma]\;
T_{u_{1}}\otimes T_{u_{2}} \,.
\label{eq:belloperatordisplace3}
\end{equation}
For Wigner-negative pure states $\sigma = \vert \psi \rangle \langle \psi \vert$ this operator is a witness of contextuality.
Whether or not it can witness nonlocality depends on the state itself.
Our search suggest that this depends, apart from the negative volume, on the `amount' of entanglement, for instance, in terms of its Schmidt coefficients.
Also note that the Bell operator here is equal to the state itself $\sigma = \mathcal{B}[\sigma]$ such that it is itself a witness for its nonlocal properties under qudit Pauli measurements.

Moreover, one can generalize this to any multipartite state,
\begin{equation}\mathcal{B}_n = \sum_{u_1,\dots,u_n \in \fld} \chi_{u_1,\dots,u_n}[\sigma] ~\bigotimes_{i =1}^{n} T_{u_i} \,,
\label{eq:bellstab2}
\end{equation}
for $n$ parties.
In analogy to the stabilizer construction with the Bell operator\,\eqref{eq:bellstab}, Wigner-negative rotated GHZ states, and presumable related ones with sufficient entanglement, lead to a nonlocal violation. Recall that the state is the probed state and defines the operator such that the quantum expectation values is always $\tr (\sigma \mathcal{B}[\sigma]) = 1$ iff $\sigma $ is a pure state.
Our search for nonlocal violations in $n=3$ also finds that the following states, which cannot be locally transform to a rotated GHZ state, lead to a Bell violation:
The fully anti-symmetric Aharonov state ($d=3$)
\begin{equation}
\vert \mathrm{A} \rangle = \frac{1}{\sqrt{6}}\left( \vert 012 \rangle + \vert 120 \rangle + \vert 201 \rangle - \vert 021\rangle - \vert 102 \rangle -\vert 210 \rangle \right) \,,
\label{eq:aharonov}
\end{equation}
leads to $B^{\max}_{\mathrm{LHV}} = 0.667$ while the quantum expectation value is $\langle \mathrm{A} \vert \mathcal{B} \vert \mathrm{A} \rangle = 1$.
And the rotated ADD-state
\begin{align}
    &\left( W_{12} \otimes \mathds{1}_3 \right) \vert \mathrm{ADD} \rangle \,,
\end{align}
with $W = \sum_{k,l \in \mathbb{Z}_3} e^{2 \iu \pi k l^2 /9} \vert kl \rangle \langle kl \vert$ and $\vert \mathrm{ADD} \rangle = \sum_{k,l\in \mathbb{Z}_3} \vert k\,l\,(k\!+\!l)\!\bmod \! 3 \rangle$.
gives maximal LHV value $B^{\max}_{\mathrm{LHV}} = 0.939$ and quantum expectation value $\langle \mathrm{ADD} \vert W_{12}^\dagger \mathcal{B} W_{12} \vert \mathrm{ADD} \rangle = 1$.

\section{Relation to other inequalities}
\label{sec:otherinequalities}

Lawrence~\cite{Lawrence2017} finds a deterministic nonlocal paradox for qutrit GHZ states shared by at least three parties each with two settings.
They use the same family of operators as Howard and Vala~\cite{Howard2012} for qutrits.

A derived Bell operator for the bipartite case with such operators is
\begin{align}
\mathcal{B}_3 &= X \otimes X + \omega \, X_{(1/3)} \otimes X_{(1/3)} + \nonumber \\ &\phantom{=+} X \otimes X_{(1/3)} + X_{(1/3)} \otimes X +\,h.c.\,, \label{eq:bellqutrit}
\end{align}
where $X_{(1/3)} = U X U^\dagger$, and $U = \mathrm{diag}(1,\omega^{2/3},\omega^{1/3})$.

Any LHV model attains
\begin{align*}
B_{3,\mathrm{LHV}}^{\max} &= \max_{a_0,a_1,b_0,b_1\in \mathbb{Z}_3} ( \omega^{a_0 + b_0} + \omega^{2a_0 + 2b_1} \\
& \hspace{2.25cm}+ \omega^{2a_1 + 2b_0} + \omega^{a_1 + b_1 + 1} +\,h.c.) \\
&= 6 + \omega + \omega^2 = 5 \,,
\end{align*}
On the other hand, the quantum expectation value, 
\[ \tr \left( \mathcal{B}_3 \, \vert \Phi \rangle \langle \Phi \vert \right) = 1 + 3 (2 \omega^{1/3} + \omega^{-2/3} ) + \,h.c.\,\approx 5.412 \,, \]
exceeds that of the classical model, $B_{3,\mathrm{LHV}}^{\max} < \tr \left( \mathcal{B}_3 \, \vert \Phi \rangle \langle \Phi \vert \right)$.

The operator $\mathcal B_3$ is a two-setting, Fourier-correlator Bell operator built from rationally rotated shift measurements 
\begin{equation}
X_{(q)}:=V_q X V_q^\dagger
\end{equation}
with rational diagonal gates
\begin{equation}\label{eq:vq_def}
V_q=\sum_{k\in\mathbb F_d}\omega^{k\{q\}}|k\rangle\langle k \vert \,,
\end{equation}
where $\{q\}\notin\mathbb Z$ is a non-character phase.
These unitaries also underpin two other known families of Bell inequalities~\cite{Collins2001,salavrakos2017bell} that we discuss in the following.

Since our formalism uses the characteristic function of states as coefficients of convenience, we will not attempt to match coefficient sequences.
Instead, we emphasize that once a small set of correlator contexts is fixed (e.g.\ the two-setting structure in Eq.\,\eqref{eq:bellqutrit} or the CGLMP/SATWAP neighbor-pair contexts), our formalism can represent the associated Bell functional by choosing coefficient operators derived from appropriately rotated Bell states (plus an irrelevant identity term and an overall rescaling, which do not affect violations).

\paragraph*{CGLMP functional.}
The Hermitian Bell operator
\begin{align}
\mathcal{B}_{\gamma} &= \sum_{k \in \fld^{\ast}} \gamma_{k} \Big( X^k_{(q_{0})} \otimes \left(X^k_{(p_{0})} + X^k_{(p_{1})} \right)\nonumber\\[-0.25cm] &\phantom{=\frac{1}{d}+ \gamma_{k} \Big(\Big(} + X^k_{(q_{1})} \otimes \left(X^k_{(p_{0})} - X^k_{(p_{1})} \right) \Big) \,, \label{eq:bellbeyond}
\end{align}
for rational numbers $q_{0},q_{1},p_{0},p_{1}$ and weights $\gamma_{k} = \gamma_{-k}$ is the discrete Fourier transform of a Collins-Gisin-Linden-Massar-Popescu (CGLMP) functional~\cite{Collins2001},
\begin{align}
\nonumber 
I_{\mathrm{CGLMP}} &= \sum_{r=0}^{d-1} \, W_r \, \Big( P(r_{00} = r) + P(r_{01} = r) \\[-0.25cm] &\hspace{1.5cm}+ P(r_{10} = r) - P( r_{11} = r)\Big)\,, \label{eq:CGLMP-prob}
\end{align}
The CGLMP choice corresponds to $(q_{0},q_{1};p_{0},p_{1})=(0,1/2;1/4,-1/4)$ and to the symmetric $\gamma_{k} :=1-2\,\min\{k,d-k\}/(d-1)$, the Fourier transform of $W_{r}$.
With these weights, the Bell operator $\mathcal{B}_{\gamma}$ satisfies the local bound
\begin{equation}
\label{eq:cglmp_lhv_value}
\max \langle I_{\mathrm{CGLMP}}\rangle_{\mathrm{LHV}} \leq 2\,.
\end{equation}
Other even weights $\gamma_{k}$ also define valid Bell functionals; the ramp $\gamma_{k}$ is distinguished by a tight, closed-form LHV bound~\cite{Collins2001}.
For the quantum values, note that $\langle\Phi \vert (X_{(q)})^{k}\otimes (X_{(p)})^{k}\,\vert \Phi\rangle = \omega^{k(p-q)}$ such that the maximally entangled state achieves
\begin{align}
\label{eq:cglmp_qvalue}
\langle\Phi \vert \mathcal{B}_{\gamma}\vert \Phi \rangle &= \sum_{k\in\fld^{\prime}}\gamma_k\Big[\omega^{k(p_{0}-q_{0})}+\omega^{k(p_{1}-q_{0})} \nonumber \\[-0.3cm] & \hspace{1.5cm} +\omega^{k(p_{0}-q_{1})}-\omega^{k(p_{1}-q_{1})}\Big] \,.
\end{align}
The freedom needed to generate the nonlocal violation with the operator~\eqref{eq:bellbeyond} using the Bell state is in the phase differences $p-q$.
Choosing the \emph{rational} offsets $(0,\tfrac12;\tfrac14,-\tfrac14)$ aligns all modes constructively in \eqref{eq:cglmp_qvalue} and lead to a violation of the Bell inequality~\eqref{eq:cglmp_lhv_value}~\cite{Collins2001}.
This shows that the non-integer phases supplied by $V_q$ control the phase of each Fourier mode $k$ via the differences $p-q$ and achieve the quantum advantage over any local model.
Taken together, these results show that diagonal unitaries with non-integer spectra $V_q$ supply the precise phase control needed to obtain a bipartite qutrit violation with non-character operators, and for any odd prime $d$, recover the CGLMP violation by aligning all Fourier modes via the phase differences.

\paragraph*{SATWAP functional.}
Salavrakos \emph{et al.} introduce a family of Bell functionals $I_{d,m}$, often referred to as the Salavrakos-Augusiak-Tura-Jordi-Wittek-Ac\'{\i}n-Pironio (SATWAP) inequalities, tailored to attain the Tsirelson bound by the maximally entangled state, for arbitrary output dimension $d$ and an arbitrary number of settings $m$ per party~\cite{salavrakos2017bell}.
In the probability form, their expression can be written as
\begin{equation}
I_{d,m}:=\sum_{k=0}^{\lfloor d/2\rfloor-1}\big(\alpha_k P_k-\beta_k Q_k\big),
\end{equation}
where
\begin{align}
&\hspace{-0.25cm}P_k:=\sum_{i=1}^m\!\Big[P({\scriptstyle{A_i=B_i+k}})+P({\scriptstyle{B_i=A_{i+1}+k}})\Big],\\
&\hspace{-0.25cm} Q_k :=\sum_{i=1}^m\!\Big[P({\scriptstyle{A_i=B_i-k-1}})+P({\scriptstyle{B_i=A_{i+1}-k-1}})\Big],
\end{align}
with all additions understood modulo $d$ and with the cyclic continuation $A_{m+1}:=A_1+1$~\cite{salavrakos2017bell}.
This form makes a distinctive feature of SATWAP explicit: it only uses the neighbor-pair contexts $(A_i,B_i)$ and $(A_{i+1},B_i)$ (a ring of $m$ links), and the links are coupled because each observable $A_i$ appears in two adjacent terms and the closure $A_{m+1}:=A_1+1$ ties the chain together.

As for the CGLMP functional, the associated optimal measurements for SATWAP are obtained from Fourier measurements conjugated by diagonal unitary operators~\cite{salavrakos2017bell}:
\begin{equation}
A_x=V_{\theta_x}^\dagger X V_{\theta_x},\qquad
B_y=V_{\vartheta_y} X V_{\vartheta_y}^\dagger,
\end{equation}
with diagonal unitary operators $V_{\theta}$ as in Eq.\,\eqref{eq:vq_def}, and phases
$\theta_x=(x-\tfrac{1}{2})/m$, $\vartheta_y=y/m$~\cite{salavrakos2017bell}.
On the maximally entangled state, the resulting probabilities (and hence the generalized correlators obtained by Fourier transform) depend on the settings only through the \emph{relative offset}
\begin{equation}
\Delta_{x,y}:=\vartheta_y-\theta_x,
\end{equation}
rather than on $x$ and $y$ separately~\cite{salavrakos2017bell}.
Because SATWAP uses only the neighbor pairs $(x,y)=(i,i)$ and $(i+1,i)$, this relative offset takes only the two values $\Delta_{i,i}=+\tfrac{1}{2m}$ and $\Delta_{i+1,i}=-\tfrac{1}{2m}$.

Thus, although the inequality is defined for $m$ settings (and its coefficients and classical bound change with $m$), in the ideal maximally entangled realization the correlator data entering $I_{d,m}$ fall into two equivalence classes, distinguished only by the sign of $\Delta$.
In our rotated-HW picture, SATWAP therefore probes two specific `relative-offset slices' of the full family of rotated correlators.
Increasing $m$ therefore does not increase the number of correlator contexts contributing to the ideal quantum value; it only changes the numerical value of the rational offset $\pm 1/(2m)$ and the weights and classical bound of the functional.
This makes the relation to our rotated-HW framework transparent.
SATWAP occupies a strict subfamily of our rotated-HW correlators, namely the $X$-line supplemented by two fixed rational-offset contexts, whereas our framework can probe the full set of $(d+1)^2$ rotated-HW contexts associated with the $d+1$ commuting subgroups.

To conclude, the CGLMP and SATWAP Bell functionals are structurally different from our construction.
Even so, their maximally-entangled-state correlations can be viewed as compatible with our framework: they arise from the same diagonal non-Clifford resources, while our construction additionally accesses correlations from a wider set of Heisenberg–Weyl contexts.
The difference in coefficient representation can be accounted for by a change of normalization and by appending contexts that yield only constant (trivial) contributions, so that the remaining terms match the characteristic-function form.

\paragraph*{Character phase functional.}
A genuine $d$-dimensional Bell functional, Eq.\,\eqref{eq:belloperatorold} appears to differ fundamentally from our construction.
Quantum implementations with rotated Pauli operators achieve a maximum violation as well~\cite{Kaniewski2018,Howard2015bell}.
However, we conjecture that the phases $\omega^{ijn}$ cannot be represented as a state's characteristic function.
While we have not proved this rigorously, all considered maps from $i\rightarrow u_i$, a phase space point, likewise $j\rightarrow u_j$ leads to a matrix $\chi_{u(i),v(i)}$ with complex eigenvalues.

\section{Unitary cube operators as non-classical resource}
\label{sec:numerics}

A particularly tractable case arises when $\rho=\vert \Phi_U \rangle\langle \Phi_U \vert$ is a Bell state rotated on one side by a diagonal unitary
\begin{equation}
U_f = \sum_{k=0}^{d-1} \omega^{f(k)} \vert k \rangle \langle k \vert\,,
\label{eq:magicunitary}
\end{equation}
with phases $u_k=\omega^{f(k)}$.
Then the relevant phase-space coefficients reduce to exponential sums over phase differences $f(k)-f(k+x)$.
If $f$ is not affine or quadratic (e.g.\ a polynomial of degree $\ge 3$ over $\fld$), Weil-type bounds imply that these sums scale at most as $\mathcal{O}(\sqrt d)$~\cite[Thm.~5.38]{Lidl1996}.
Consequently, one obtains dimension-dependent analytic bounds on the non-contextual value (and thus on contextuality witnesses derived from Pauli measurements), as worked out for unitary cube operators in Appendix~\ref{app:wignercube}.
In contrast, for the LHV bound we rely on the discrete optimization~\eqref{eq:bellineq}.

As a concrete non-Clifford family we use the qudit analogues of the qubit $\pi/8$-gate introduced by Howard and Vala~\cite{Howard2012} (see also~\cite{Howard2015bell}).
They start from diagonal unitaries~\eqref{eq:magicunitary} and choose $f$ such that conjugation maps Pauli operators to Clifford operators (i.e.\ $U_f$ lies in the third level of the Clifford hierarchy). 
For odd primes $d>3$ this yields cubic phase functions.
In the following, we refer to these generalized $\pi/8$-gates as \emph{unitary cube operators} and denote them by $U_\nu$.

For completeness, we recall the explicit parametrization used in~\cite{Howard2012,Howard2015bell} (see also Appleby's Clifford parametrization~\cite{appleby2005symmetric,Appleby2009}).
For $d>3$ one may take
\[
f(k) = \nu_k = 12^{-1} k \bigl(\gamma + k (6z + (2k+3)\gamma)\bigr) + \epsilon k
\]
with $z,\epsilon\in\fld$ and $\gamma\in\fld^\ast$.
such that
\[
U^{\phantom{\dagger}}_\nu X U^\dagger_\nu 
= \omega^\epsilon X Z^z \sum_{k \in \fld} \omega^{2^{-1}\gamma k^2} \vert k \rangle \langle k \vert\,.
\]

The qutrit case differs from odd primes $d>3$: over $\mathbb F_3$ one has $k^3\equiv k$, so any would-be `cubic' phase polynomial reduces to at most quadratic order and therefore generates only Clifford operations.
In particular, the unitary cube-operator approach based on (integer-valued) character polynomials cannot produce genuinely non-Clifford phases in $d=3$.
Therefore, it cannot access the qutrit resources needed for stronger contextuality and nonlocality witnesses.

Howard and Vala~\cite{Howard2012} resolve this by going beyond character spectra and introducing \emph{fractional} phases (fixed branches of roots of unity).
Concretely, they take a third root of the characters and define the `magic' diagonal unitary via fractional exponents, e.g.\ by choosing $\nu_k=(6zk^2+2\gamma k+3k\epsilon)/3$ in Eq.~\eqref{eq:magicunitary}.
As discussed in Sec.\:\ref{sec:otherinequalities}, other known inequalities have used such operators with spectra beyond the qudit characters to minimize the number of local settings and generalize qubit inequalities such as the CHSH~\cite{chsh1969} and the Mermin GHZ paradox~\cite{mermin1990simple}.

\paragraph*{Wigner negativity of rotated Bell states}

For $d>3$ the Wigner function admits an explicit cubic character-sum form (derived in Appendix~\ref{app:wignercube}),
\begin{equation}
W^{\nu}_{u_1,u_2}
=\frac{1}{d^3}\,\delta_{(u_1)_x,(u_2)_x}
\sum^{d-1}_{k=0} \omega^{ a_3 k^3 + a_1 k}\,,
\label{eq:wignerneg}
\end{equation}
where $a_1=a_1(u_1,u_2)$ and $a_3$ depend on the unitary parameters.
Concretely,
$a_1(u_1,u_2) = \epsilon + (u_1)_z + (u_2)_z + z (u_1)_x + 2^{-1}\gamma \bigl((u_1)_x^2 - (u_1)_x + 6^{-1}\bigr)$
and $a_3 = 24^{-1}\gamma$.
Thus, as $(u_1,u_2)$ vary subject to $(u_1)_x=(u_2)_x$, the coefficient $a_1$ ranges over $\fld$ while $a_3$ ranges over $\fld^\ast$.
Cubic character sums are difficult to analyze in full generality, but a brute-force search over $(a_1,a_3)$ readily finds negative values of~\eqref{eq:wignerneg} for all $d>3$.
This negativity is pivotal for achieving contextuality and Bell violations with the Bell functional defined by $\vert\Phi_\nu\rangle$.

\paragraph*{Non-contextual and Bell bounds}
The non-contextual bound is governed by the maximal phase-space value of the Wigner function.
In Appendix~\ref{app:wignercube} we derive the analytic estimate (cf.\ Eq.\,\eqref{eq:cubewignerbound}), for $d>3$:
\begin{equation}
B_{\mathrm{NC}}^{\max} \leq \frac{2}{\sqrt{d}}\,.
\end{equation}
For $d=3$, an exhaustive search yields $B_{\mathrm{NC}}^{\max} < 0.844$.
Table~\ref{tab:charactersumvalue} lists the resulting values $d^2 \max_{u_1,u_2\in\fld^2} W^\nu_{u_1,u_2}$ (as well as further extremal quantities) for odd primes $d\le 23$.

To assess nonlocality, we evaluate the LHV optimization~\eqref{eq:bellineq} for $\mathcal{B}_\nu$ (equivalently $\mathcal{B}[\Phi_\nu]$), which yields tight values of $B^{\max}_{\mathrm{LHV}}$ for small odd prime $d$. 
The results are reported in Table~\ref{tab:charactersumvalue}; empirically, $B^{\max}_{\mathrm{LHV}}$ decreases with increasing $d$ for the unitary cube family.

\begin{table}[t]
\centering
\hspace*{-0.5cm}
\begin{tabular}{Sl||Sr|Sr|Sr|Sr||Sr} %
$d$ & $N^{\nu}_{\max}$ & $d^3 W^{\nu}_{\min}$ & $C_{\min} $ & \vspace{0.05cm} $ B^{\max}_{\mathrm{NC}} $ & $B^{\max}_{\mathrm{LHV}}$ \vspace{0.05cm} \\ \cline{1-6}
$2^{\ast\ast}$ & $0.354$ & $-0.707$ & $-\phantom{1}0.707$ & $0.213$ & $0.213^{\phantom{\dagger}}$\\
\cdashline{1-6} $3^\ast$ & $0.293$ & $ -0.879$ & $-\phantom{1}0.879$ & $0.844$ & $0.960^{\phantom{\dagger}}$ \\
5 & $0.447$ & $-2.236$ & $-\phantom{1}2.236$ & $0.724$ & $0.877^{\phantom{\dagger}}$\\
7 & $0.725$ & $-4.406$ & $-\phantom{1}4.406$ & $0.677$ & $0.829^{\phantom{\dagger}}$\\
\cdashline{1-6} 11 & $0.914$ & $-4.211$ & $-\phantom{1}8.595$ & $0.535$ & $0.769^{\phantom{\dagger}}$ \\
13 & $1.102$ & $-6.953$ & $-10.651$ & $0.442$ & $0.758^{\phantom{\dagger}}$\\
17 & $1.251$ & $-7.030$ & $-14.728$ & $0.437$ & \vspace{0.5cm} $0.737^\dagger$ \\
19 & $1.437$ & $-6.438$ & $-16.755$ & $0.449$ & $0.718^\dagger$\\
23 & $1.531$ & $-8.654$ & $-20.795$ & $0.371$ & $0.674^\dagger$
\end{tabular}
\caption{Extremal values of the Wigner function $W^\nu_{u_1,u_2}$ in Eq.\,\eqref{eq:wignerneg}, maximal values $N^{\nu}_{\max}= \max_\nu N^{\nu}$ of its negativity volume $N^{\nu} := N[ \Phi_\nu ]$ and the non-contextual bound $B^{\max}_{\mathrm{NC}}$ from Eq.\,\eqref{eq:ncineq}, minimal values $ W^{\nu}_{\min} = \min_{u_1,u_2\in \fld^2} W^{\nu}_{u_1,u_2}$, as well as values of $C_{\min}$ (Eq.\,\eqref{eq:minchi}) for small prime dimension $d$. All values are the result of a brute-force numerical search.
For $d \leq 7$, we saturate the bound of Eq.\,\eqref{eq:degreebound} and achieve $C_{\min} = d^3\min W^{\nu}_{v_1,v_2}$.
~* For $d=3$, we instead use $ \chi (f) = 1 + 2 \cos \left(8 \pi/9 \right) $ as we allow for cube roots of the character.
~** For $d=2$, we use the Wigner function adapted to qubits for the Bell state and the $\pi/8$-gate in Eq.\,\eqref{eq:appqubitwigner}. Its minimum value is $-1/\sqrt{2} \approx -0.707$, corresponding to Tsirelson's bound~\cite{Cirelson1980}.
Note that the Bell operator\,\eqref{eq:belloperatorgen} and analysis thereof is not apt to qubits.
We find the nonlocal bound $B^{\max}_{\mathrm{LHV}}$ from Eq.\,\eqref{eq:bellineq} via a brute force search until and including $d=13$.
~$\dagger$ for $d=17,19,23$, we use a sound but heuristic optimization algorithm, see Appendix\:\ref{app:numerics}.}
\label{tab:charactersumvalue}
\end{table}

\section{Conclusion and outlook.}
We constructed qudit Bell inequalities in all odd prime dimensions that are designed around the phase-space structure of qudit Pauli measurements.
Our starting point is that, in odd dimensions, Wigner negativity is equivalent to contextuality for projective Pauli measurements~\cite{Gross2006,Howard2014,Delfosse2017}, and hence it is a necessary resource for Bell nonlocality in this setting.
Accordingly, we introduced a family of Bell operators whose coefficients are given by the characteristic function of a rotated entangled state. 
For bipartite systems this yields a CHSH-type generalization in which the target state attains the algebraic value, while the classical bounds are controlled by phase-space constraints. 
In particular, the expectation value of our Bell operator can be interpreted as a singlet-fraction witness and is bounded in terms of the maximal Wigner value and the negativity volume.

As a concrete example family, we specialized to the diagonal non-Clifford unitaries of Howard and Vala~\cite{Howard2012} (unitary cube operators) that rotate stabilizer states into Wigner-negative states while mapping Pauli operators into the Clifford group~\cite{appleby2005symmetric,Appleby2009}. 
For these operators we obtain analytic control of the non-contextual bound and compute tight LHV bounds numerically in small prime dimensions. 
Moreover, we showed that the same construction extends naturally to multipartite stabilizer states, where stabilizer symmetries and the diagonal structure of the rotation can be exploited to reduce the number of required measurement settings.

Beyond character spectra, we also identified diagonal unitaries with rational, non-character eigenphases as an additional resource. 
These rotations introduce phase offsets that coherently align correlation modes for maximally entangled states, while deterministic local strategies---restricted to character-valued phases---cannot reproduce this sensitivity. 
Finally, we related our construction to the CGLMP inequality and to the SATWAP family~\cite{salavrakos2017bell}. 
Although these functionals are structurally different from our characteristic-function-based Bell operators, their optimal maximally entangled correlations are generated by the same class of diagonal rotations in the HW framework.
In particular, SATWAP probes only neighbor-pair contexts and therefore reduces, in the ideal realization, to two `relative-offset' correlator classes; in our language it occupies a strict subfamily of the rotated-HW correlators.
Bridging the gap from these standard probability-form coefficients to characteristic-function coefficients amounts to rescaling and adding trivial contexts, which do not change the presence of a violation.

Several open directions remain. 
On the theory side, it would be interesting to obtain sharper analytic LHV bounds for broader classes of rotations and to understand systematically when the characteristic-function construction yields near-optimal Bell inequalities. 
On the implementation side, realizing the required non-Clifford operations (including rational-phase diagonals) depends strongly on the experimental platform; possible candidates include superconducting circuits, cold atoms, and optical systems (e.g.\ involving Gottesman--Kitaev--Preskill states)~\cite{PhysRevA.64.012310}.

\paragraph*{Acknowledgements.} We sincerely thank Mark Howard for valuable insights. We acknowledge funding from Horizon Europe Research and Innovation
Actions under Grant Agreement 101080173 CLUSTEC and from the Plan France 2030 through the ANR-22-PETQ-0006NISQ2LSQ and ANR-22-PETQ-0007 EPiQ projects.

\bibliography{main}

\onecolumngrid
\pagebreak
\appendix

\section{Wigner Negativity}
\label{app:wignercube}

In the following, we discuss some technical features of the Wigner function $ W^{\nu}_{u_1,u_2}$ and optimizations of its Wigner negativity with different polynomials $f$ in Eq.\,\eqref{eq:magicunitary}.
The polynomial $\nu_k$ is proportional to the third Dickson polynomial~\cite{Dickson1896}, whose value set has been studied in~\cite{Chou1988}.
From~\cite[Weil's Theorem 5.38, p.\@ 223]{Lidl1996}, the value of the corresponding character sum is bound by $2\sqrt{d}$ such that
\begin{equation} \label{eq:cubewignerbound}
   \vert W^{\nu}_{u_1,u_2} \vert \leq \frac{2}{d^2\sqrt{d}} \,.
\end{equation}
The variance of the character sum with Dickson polynomials over uniformly random values $a_3/a_1^3$ if $a_1 \neq 0$, otherwise over values of $a_3$, is of order $O(\sqrt{d})$~\cite[Lemma 2]{Ma2019}, while its mean value is of order $O(1)$.
Table~\ref{tab:charactersumvalue} lists minimum values of $W^{\nu}_{v_1,v_2}$ for small $d$.

To find unitary operators that achieve larger Wigner negativity, consider diagonal operators from Eq.\,\eqref{eq:magicunitary} for some polynomial $f$ with $3 < \mathrm{deg} (f) < d$.
From Eq.\,\eqref{eq:appwignergen}, the Wigner function of the Bell state under the adjoint action of $U$ is proportional to \( \chi (f) = \sum_{k=0}^{d-1} \omega^{g_k} \,,\) for the polynomial $g_{k} \sim f_{k} - f_{-k}$.
The lower bound of $\chi(g)$ is \begin{equation} \min_{g} \chi (g) \geq C_{\min}= 1 - (d-1) \cos \left( \pi/d \right) \label{eq:minchi} \end{equation} since $g_0 =0$ and $a = (d\pm 1)/2$ minimizes $\Re (\omega^a)$.
For comparison, numerical values of $\min_{g} \chi(g)$ for small odd prime numbers $d$ are in Tab.\,\ref{tab:charactersumvalue} which differ from $W_{\min}= \min_{u_1,u_2 \in \fld^2} W^{\nu}_{u_1,u_2}$ for all $d>7$.
For $d \rightarrow \infty$, $C_{\mathrm{min}} \rightarrow 2-d$ while $W_{\min} \geq -2/d^2\sqrt{d}$ from Eq.\,\eqref{eq:cubewignerbound}.
To estimate if $C_\mathrm{min}$ is achievable by a polynomial $g$, we study the value set $g(\fld) $ of a polynomial over finite fields.
Specifically, $C_\mathrm{min}$ requires $g$ to have a value set of three elements.
The value set of any polynomial has the lower bound \begin{equation}\left\lceil \frac{d-1}{\mathrm{deg}(g)} \right\rceil +1 \leq \vert g(\fld) \vert \,, \label{eq:degreebound}\end{equation} such that, for $\vert g(\fld) \vert = 3$ to achieve the $C_{\min}$ in Eq.\,\eqref{eq:minchi}, it is necessary that $ \mathrm{deg}(g) \geq (d-1)/2$,~\cite{Wan1993}.

We evaluate the Wigner function of the state $\vert \Phi_{f} \rangle = U_{f} \vert \Phi \rangle$ for a $U_{f}$ from Eq.\,\eqref{eq:magicunitary} with a polynomial $f(k) = f_k$,
\begin{align}
 W_{u_1,u_2} (\vert \Phi_{f} \rangle \langle \Phi_{f} \vert)
 &= \frac{1}{d^2} \tr \left( A_{u_1} A_{u_2} \vert \Phi_{f} \rangle \langle \Phi_{f} \vert \right) \nonumber\\
 &= \frac{1}{d^5} \hspace{-0.2cm}\sum_{\substack{j,k \in \fld \\ v_1, v_2 \in \fld^2}} \hspace{-0.2cm} \omega^{[u_1,v_1] + [u_2,v_2]} \langle j j \vert U^{\phantom{\dagger}}_{{f}} \, T_{v_1} U^\dagger_{{f}} \otimes T_{v_2} \vert k k \rangle \nonumber \\
 &= \frac{1}{d^5} \hspace{-0.2cm}\sum_{\substack{ j,k \in \fld \\v_1, v_2 \in \fld^2} } \hspace{-0.2cm}\omega^{[u_1,v_1] + [u_2,v_2] + k((v_1)_z +(v_2)_z) + 2^{-1}((v_1)_x (v_1)_z + (v_2)_x (v_2)_z) } \langle j j \vert U^{\phantom{\dagger}}_{{f}} \, X^{(v_1)_x} U^\dagger_{{f}} \otimes X^{(v_2)_x} \vert k k \rangle \nonumber \\
 &= \frac{1}{d^5} \hspace{-0.2cm}\sum_{\substack{ k \in \fld \\ v_1, v_2 \in \fld^2}} \hspace{-0.2cm}\delta_{(v_1)_x,(v_2)_x} \,\omega^{[u_1,v_1] + [u_2,v_2] + k((v_1)_z +(v_2)_z) + 2^{-1}((v_1)_x (v_1)_z + (v_2)_x (v_2)_z)+{f}_{k+(v_1)_x} - {f}_{k^{\phantom{(1)}}} } \nonumber\\
 ^{(v_x :=(v_1)_x)}&= \frac{1}{d^5} \hspace{-0.2cm} \sum_{\substack{ v_x,k \in \fld \\ (v_1)_z, (v_2)_z \in \fld}} \hspace{-0.2cm} \omega^{((u_1)_z + (u_2)_z)v_x - (u_1)_x (v_1)_z - (u_2)_x (v_2)_z + k( (v_1)_z + (v_2)_z) + 2^{-1}v_x( (v_1)_z + (v_2)_z) + {f}_{k+v_x} - {f}_k } \nonumber \\
 &= \frac{1}{d^3} \sum_{v_x,k \in \fld} \delta_{k,(u_1)_x - 2^{-1} v_x} \delta_{k,(u_2)_x - 2^{-1} v_x} \omega^{((u_1)_z + (u_2)_z)v_x +{f}_{k+v_x} - {f}_k } \nonumber \\
 &= \frac{1}{d^3} \delta_{(u_1)_x, (u_2)_x} \sum_{v_x \in \fld} \omega^{((u_1)_z + (u_2)_z)v_x +{f}_{(u_1)_x + 2^{-1} v_x} - {f}_{(u_1)_x - 2^{-1} v_x} }\,.\label{eq:appwignergen}
\end{align}

For any bipartite state $\rho$, the Bell operator's expectation value from Eq.\,\eqref{eq:belloperatorgen} is
\begin{align}
    \tr \left( \mathcal{B}_{\nu} \rho \right) &= \sum_{u_{1},u_{2} \in \fld^2} W_{u_{1},u_{2}} (\vert \Phi_{\nu} \rangle \langle \Phi_{\nu} \vert) \tr \left( U^{\phantom{\dagger}}_{\nu} A_{u_1} U^{\dagger}_{\nu} \otimes A_{u_2}\rho \right) \\
    &= \tr \left( \left( U^{\phantom{\dagger}}_{\nu} \otimes \mathds{1} \right) \vert \Phi_{\nu} \rangle \langle \Phi_{\nu} \vert \left( U^{\dagger}_{\nu} \otimes \mathds{1} \right) \rho \right) \\
    & \langle \Phi \vert \rho \vert \Phi \rangle
    \label{eq:appsingletfraction}
\end{align}

For $d= 3$, $\nu_k = (6zk^2 + 2\gamma k + 3k\epsilon)/3$ and
\begin{equation}
    W_{u_1,u_2} (\vert \Phi_\nu \rangle \langle \Phi_\nu \vert) 
    =\frac{1}{d^3} \delta_{(u_1)_x,(u_2)_x} \left( 1 + \omega^{a_1((u_1)_z,(u_2)_z) - a_3/3 } + \omega^{-a_1((u_1)_z,(u_2)_z) + a_3/3 } \right) \,,
\end{equation}
with $a_1((u_1)_z,(u_2)_z) = (u_1)_z + (u_2)_z + z (u_1)_x + \epsilon$ and $a_3 = \gamma$.

\section{Relation to the CHSH inequality}
\label{app:chsh}

For qubits ($d=2$) an equivalent quantity can be characterized with the $T$-gate $T = \vert 0 \rangle \langle 0 \vert + e^{\iu \pi/4} \vert 1 \rangle \langle 1 \vert $, such that
\begin{align}
    W_{u_1,u_2}( (T \otimes \mathds{1})\vert \Phi \rangle \langle \Phi \vert(T^\dagger \otimes \mathds{1}) ) &= \frac{1}{2^4} \sum_{\substack{(v_1)_x,(v_1)_z, \\ (v_2)_x,(v_2)_z = 0,1 }} (-1)^{(u_1)_z (v_1)_x + (u_1)_x (v_1)_z + (u_2)_z (v_2)_x + (u_2)_x (v_2)_z} \nonumber\\[-0.25cm] & \hspace{3cm} \cdot \tr ( T^\dagger \sigma_{((v_1)_x,(v_1)_z)} T \otimes\sigma_{((v_2)_x,(v_2)_z)} \vert \Phi \rangle \langle \Phi \vert) \nonumber \\
    &= \frac{1}{2^4} \Big( 1 + (-1)^{(u_1)_x + (u_2)_x} \nonumber\\ &\phantom{=\frac{1}{2^4}} + (-1)^{(u_1)_z + (u_2)_z} \big( 1-(-1)^{(u_1)_x + (u_2)_x} + (-1)^{(u_1)_x} + (-1)^{(u_2)_x } \big)/\sqrt{2} \Big) \,, \label{eq:appqubitwigner}
\end{align}
with the Pauli operators $\sigma_{(0,0)} = \mathds{1}$, $\sigma_{(1,0)} = \sigma_x$, $\sigma_{(1,1)} = \sigma_y$, $\sigma_{(0,1)} = \sigma_z$.
The evaluations of the expectation values and nonlocal violations for the qudit Bell operator in Eq.\,\eqref{eq:belloperatordisplace} do not straightforwardly generalize to qubits using the above Wigner function.
The reason is that the assignments of an LHV model are different from qudit Pauli operators to qubit Pauli operators if one measures a complete set of operators generating mutually unbiased bases.
However, in case where $U_{\nu} = T$, the equivalent Bell operator for qubits in Eq.\,\eqref{eq:belloperatordisplace} is
\begin{equation}
    \mathcal{B}^{\phantom{\prime}}_{2} = \mathds{1}\otimes \mathds{1} + \sigma_z \otimes \sigma_z - \frac{1}{\sqrt{2}} \Big( T \sigma_y T^\dagger \otimes \sigma_y + T \sigma_x T^\dagger \otimes \sigma_y + T \sigma_y T^\dagger \otimes \sigma_x - T \sigma_x T^\dagger \otimes \sigma_x \Big)\,,\label{eq:qubitchshtgate}
\end{equation}
leading to an inequality $ 1 + A_2 B_2 - \left( A_0 B_0 + A_1 B_0 + A_0 B_1 - A_1 B_1 \right)/\sqrt{2} \leq 2 + \sqrt{2}$ that is analogous to the Clauser-Horne-Shimony-Holt (CHSH) $ A_0 B_0 + A_1 B_0 + A_0 B_1 - A_1 B_1 \leq 2$.

\section{Bounds on Gross' Wigner function}

We consider a general pure bipartite qudit state $\vert \Psi \rangle$.
The sum over absolute values can be expressed by the negative sum of all negative values, the negativity volume $N[\Psi]$, with $\sum_{v_1,v_2\in \fld^{2}} \vert W_{v_1,v_2}[\Psi] \vert = 1+2 N[\Psi]$.
H\"{o}lder's inequality~\cite{holder1889mittelwerthssatz} bounds the negativity volume through
\begin{equation}
    \Bigg( \sum_{v_1,v_2 \in \fld^{2}} \left\vert W_{v_1,v_2}[\Psi] \right\vert \,\Bigg)^2 \leq d^{4} \sum_{v_1,v_2 \in \fld^{2}} W_{v_1,v_2}[\Psi]^2 = d^{2} \,.
\end{equation}
Altogether, it is
\begin{equation} 
\label{eq:wignerbounds}
    \frac{1}{d^{3}} \leq \frac{1}{d^{2}\left( 1+2 N[\Psi] \right) } \leq \max_{v_1,v_2\in \fld^{2}} \vert W_{v_1,v_2}[\Psi] \vert \leq \frac{1}{d} \,,
\end{equation}
for any pure bipartite state $\vert \Psi \rangle$.

Lastly, for qudits, we derive Eq.\,\eqref{eq:volumebound} from
\begin{align}
    \tr \left( \rho \mathcal{B} \right) &= d^2 \sum_{u_1,u_2 \in \fld^2} W^\nu_{u_1,u_2} W_{u_1,u_2} \left( (U_{\nu} \otimes \mathds{1})^\dagger \rho \, (U_{\nu} \otimes \mathds{1}) \right) \nonumber \\
    &\leq d^2 \max_{u_1,u_2 \in \fld^2} \vert W^\nu_{u_1,u_2} \vert \sum_{u_1,u_2 \in \fld^2}  \vert W_{u_1,u_2} \left( (U_{\nu} \otimes \mathds{1})^\dagger \rho \, ( U_{\nu} \otimes \mathds{1}) \right) \vert \,. \label{eq:appcalcvolumebound} 
\end{align}

\section{Lemma (relative-offset reduction on the Bell state).}

\label{lem:rel_offset}
Let $|\Phi\rangle=\sum_{k=0}^{d-1}|k\,k\rangle/\sqrt d$ and define the diagonal offset unitaries
\[
V_q \;=\; \sum_{k\in\fld}\omega^{k q}\,|k\rangle\langle k|,\qquad
X_{(q)} \;:=\; V_q X V_q^\dagger,
\]
for (possibly non-integer) rational $q$. Then for any $q_x,q_y\in\mathbb Q$ and any $l\in\fld$,
\begin{equation}
\label{eq:rel_offset_reduction}
    \big\langle \Phi \big|\, X_{(q_x)}^{\,l}\otimes X_{(q_y)}^{\,-l}\,\big|\Phi\big\rangle
    \;=\;
    \big\langle \Phi \big|\, X^{l}\otimes X_{(q_y-q_x)}^{\,-l}\,\big|\Phi\big\rangle,
\end{equation}
and hence this correlator depends on $(q_x,q_y)$ only through the difference $q_y-q_x$.

\emph{Derivation.}
Using $\langle \Phi|A\otimes B|\Phi\rangle=\frac{1}{d}\tr(A^{\mathsf T}B)$, the fact that $V_q^{\mathsf T}=V_q$ (diagonal) and $X^{\mathsf T}=X^{-1}$, we obtain
\[
\begin{aligned}
\langle \Phi|X_{(q_x)}^{\,l}\otimes X_{(q_y)}^{\,-l}|\Phi\rangle
&=\frac1d\tr\!\left(\big(V_{q_x}X^lV_{q_x}^\dagger\big)^{\mathsf T}\,V_{q_y}X^{-l}V_{q_y}^\dagger\right)\\
&=\frac1d\tr\!\left(V_{q_x}X^{-l}V_{q_x}^\dagger\,V_{q_y}X^{-l}V_{q_y}^\dagger\right)\\
&=\frac1d\tr\!\left(X^{-l}\,V_{q_y-q_x}\,X^{-l}\,V_{q_y-q_x}^\dagger\right)\\
&=\langle \Phi|X^{l}\otimes \big(V_{q_y-q_x}X^{-l}V_{q_y-q_x}^\dagger\big)|\Phi\rangle,
\end{aligned}
\]
which is Eq.~\eqref{eq:rel_offset_reduction}.

\paragraph*{Corollary (minimal settings for the SATWAP correlator subfamily).}
\label{cor:min_settings}
Consider the SATWAP optimal offsets $q_x=(x-\tfrac12)/m$ and $q_y=y/m$. The SATWAP Bell value involves only the pairs $(x,y)=(i,i)$ and $(x,y)=(i+1,i)$, for which
\[
q_y-q_x=\frac{y}{m}-\frac{y-\tfrac12}{m}=+\frac{1}{2m},\qquad
q_y-q_{x}=\frac{y}{m}-\frac{y+\tfrac12}{m}=-\frac{1}{2m}.
\]
Hence, on $|\Phi\rangle$, all correlators entering the SATWAP value reduce to the two relative-offset families
$\langle X^{l}\otimes X_{(+1/2m)}^{-l}\rangle_\Phi$ and $\langle X^{l}\otimes X_{(-1/2m)}^{-l}\rangle_\Phi$ (for all $l\in\fld^\ast$), which can be obtained from only two correlation contexts.

If one requires at least two settings on each side, let $\delta:=1/(4m)$ and define two local settings per party
\[
A_0:=X_{(+\delta)},\quad A_1:=X_{(-\delta)},\qquad
B_0:=X_{(-\delta)},\quad B_1:=X_{(+\delta)}.
\]
Then the two contexts $(A_0,B_0)$ and $(A_1,B_1)$ realize the required relative offsets:
$q(B_0)-q(A_0)=-2\delta=-1/(2m)$ and $q(B_1)-q(A_1)=+2\delta=+1/(2m)$.
By Lemma~\ref{lem:rel_offset}, the corresponding outcome statistics determine all Fourier-mode correlators
$\langle A_x^l\,\bar B_y^{\,l}\rangle_\Phi$ entering the SATWAP value.

If one is only interested in reproducing the correlators entering the SATWAP score on $|\Phi\rangle$ (rather than implementing the full $m$-setting Bell scenario), an even more economical correlator-realization uses one $X$-setting on Alice and two settings $X_{(\pm 1/2m)}$ on Bob.

\section{Numerical evaluation of the LHV bound}

\label{app:numerics}

To obtain the local-hidden-variable (LHV) bound $B^{\max}_{\mathrm{LHV}}$ we maximize the Bell functional over deterministic local strategies.
In our implementation, a deterministic strategy is parametrized by one value per measurement setting (one per commuting Pauli subgroup), and the predicted phase for a label $(x,z)$ with $x\neq 0$ depends only on the corresponding slope $k=z/x$. 
The contribution of the remaining $x=0$ subgroup is treated separately and added back as a constant $+1/d$ after the optimization. 

The resulting discrete maximization is carried out by an alternating best-response heuristic: starting from a random initialization on Alice’s side, we compute Bob’s best response exactly, then Alice’s best response exactly, and iterate until convergence. 
To reduce the risk of local optima, we repeat this procedure for many random restarts (and a small set of fixed initial seeds) and keep the best value found. 

Efficiency comes from exploiting the structure of the coefficient array for unitary-cube instances: the action of the coefficients on a phase table can be evaluated using FFT-based convolutions instead of explicitly forming a dense matrix, and each best-response step can also be computed via FFTs along the slope lines. 
Finally, when scanning over unitary parameters (e.g.\ $\gamma\in\fld^\ast$ and $z\in\fld$), we use an adaptive procedure: each instance is first evaluated with a smaller “probe” restart budget, and only promising instances are refined with the full budget.

\paragraph*{Pseudocode.}
\begin{quote}
\textbf{Algorithm A1 (LHV local search for fixed instance)} \\
\textit{Input:} prime $d$, precomputed FFT kernels $\mathrm{FFT}[c_x(\cdot)]$ for $x\in\fld^\ast$, restart count $R$, max iters $T$. \\
\textit{Output:} best value estimate $V^\star \approx B_{\mathrm{LHV}}^{\max}$. \\[2pt]
1.\ Initialize $V^\star\leftarrow -\infty$. \\
2.\ For each restart $r=1,\dots,R$: \\
\hspace*{1.5em}2.1 Sample initial slope parameters $a^{(0)}\in\fld^d$ (plus optional fixed seeds). \\
\hspace*{1.5em}2.2 For $t=0,\dots,T-1$: \\
\hspace*{3em}(i) Build phase matrix $p_A[x,z]=\omega^{x\,a^{(t)}_{z/x}}$ for $x\neq 0$. \\
\hspace*{3em}(ii) Compute $v = \chi\,p_A$ using FFT-based convolution per $x$. \\
\hspace*{3em}(iii) Compute exact best response $b^{(t+1)}=\mathrm{BestResponse}(v)$ by slope-wise FFTs. \\
\hspace*{3em}(iv) Build $p_B[x,z]=\omega^{x\,b^{(t+1)}_{z/x}}$ and compute $w=\chi\,p_B$. \\
\hspace*{3em}(v) Update $a^{(t+1)}=\mathrm{BestResponse}(w)$ by slope-wise FFTs. \\
\hspace*{3em}(vi) Stop early if $(a^{(t+1)},b^{(t+1)})$ is unchanged. \\
\hspace*{1.5em}2.3 Evaluate final score $V_r=\Re(\langle p_B,\chi p_A\rangle)/d^2 + 1/d$ and set $V^\star\leftarrow\max(V^\star,V_r)$. \\
3.\ Return $V^\star$.

\vspace{0.8em}
\textbf{Algorithm A2 (Adaptive scan over unitary-cube parameters)} \\
\textit{Input:} parameter grid (e.g.\ $\gamma\in\fld^\ast$, $z\in\fld$), budgets $R_{\mathrm{probe}}\ll R_{\mathrm{full}}$. \\
\textit{Output:} best instance (minimizer or maximizer as needed) and its estimated value. \\[2pt]
1.\ Evaluate the first instance with $R_{\mathrm{full}}$ restarts and store the current best value. \\
2.\ For each remaining instance: \\
\hspace*{1.5em}2.1 Run Algorithm A1 with $R_{\mathrm{probe}}$ restarts to obtain a probe value. \\
\hspace*{1.5em}2.2 If the probe cannot beat the current best (within tolerance), discard. \\
\hspace*{1.5em}2.3 Otherwise, continue the same RNG stream and extend to $R_{\mathrm{full}}$ restarts; update the best if improved. \\
3.\ Return the best instance and value.
\end{quote}

\end{document}